\documentclass[12pt]{iopart}

\usepackage{graphicx}
\usepackage{color}
\begin{document}

\title{Inferring network topology via propagation process}

\author{An Zeng\footnote{an.zeng@unifr.ch}}

\address{ Department of Physics, University of Fribourg, Chemin du Mus\'{e}e 3, CH-1700 Fribourg, Switzerland\\
}

\begin{abstract}

Inferring the network topology from the dynamics is a fundamental problem with wide applications in geology, biology and even counter-terrorism. Based on the propagation process, we present a simple method to uncover the network topology. The numerical simulation on artificial networks shows that our method enjoys a high accuracy in inferring the network topology. We find the infection rate in the propagation process significantly influences the accuracy, and each network is corresponding to an optimal infection rate. Moreover, the method generally works better in large networks. These finding are confirmed in both real social and nonsocial networks. Finally, the method is extended to directed networks and a similarity measure specific for directed networks is designed.
\end{abstract}
\maketitle

\section{Introduction.}

Spreading processes widely exist in various fields including physics, chemistry, medical science, biology and sociology~\cite{RMP801275}. For example, reaction diffusion processes~\cite{NP3276}, pandemics~\cite{PRL863200}, cascading failures in electric power grids~\cite{PRL93098701} and information dissemination~\cite{EPL8838005} can be naturally described by the framework of spreading. In the past decade, spreading on complex networks has been intensively studied. Studies have revealed that the spreading results is strongly influenced by the
network topologies~\cite{PRL89108701,PRL90028701,PRL97088701,PRL105218701}. With these understanding, some network manipulating methods are designed to hinder spreading in the case of diseases or accelerate spreading in the case of information dissemination~\cite{EPL9518005}.

Recently, more and more attention has been paid to the microscopic level when studying the spreading process on networks~\cite{PLA3771031}. Since the local structure around each node can be very different, the final spreading coverage varies from several nodes to the entire network when the propagation originates from distinct nodes. So far, many methods, such as the k-shell~\cite{NP6888} and the leaderrank~\cite{Plos21202}, have been proposed to rank the spreading ability of the nodes (i.e., how many nodes will finally be reached when the spreading originates from this single node).

A fundamental problem related to the spreading process is how to infer the network topology from the observation of the spreading results. If this question is answered, we could, for instance, have a better understanding of the organization of the terrorists (social networks) and the structure of some biology systems (metabolic networks). Since building the relation between the dynamics and network structure is a crucial problem, much effort has been made in this direction~\cite{PR424175}. In ref.~\cite{NJP13013004}, the authors design a method to reconstruct the network based on the observation of some oscillation taking place on networks. Moreover, noise is found to lead to a general, one-to-one correspondence between the dynamical correlation and the network connections~\cite{PRL104058701}. Very recently, the oscillation is also used to predict the missing nodes in network~\cite{PRE85065201}. Even though the spreading process widely exist in many real systems, so far little has been investigated in the literature about inferring network topology based on the spreading. The closest studies are ref.~\cite{PRE84056105,PRL109068702} where the spreading results are used to identify the initial spreader of certain disease or information.

In this paper, we proposed a simple method to uncover the network topology. The basic idea is that the similarity between nodes can be estimated based on the spreading results. We test our method in two well-known artificial network models. The results shows that our method has a high accuracy in inferring network topology. Moreover, the infection rate of the spreading is found to significantly influence the inferring accuracy and each network has an optimal infection rate. We also validate our method in both real social and nonsocial networks. Finally, we design a new similarity measure and extend our method to directed networks. The new similarity measure is shown to remarkably improve the inferring accuracy compared to the existing similarity measures.

\section{Model.} We consider a network with $N$ nodes and $E$ links. The network is represented by an adjacency matrix $A$, where $a_{ij}=1$ if there is a link between node $i$ and $j$, and $a_{ij}=0$ otherwise. To simulate the spreading process on networks, we employ the SIR model~\cite{RMP801275}. Actually, this model has been used to simulate many different propagation process. Without losing any generality, we consider the online information spreading as an example in this paper. We assume that each user has probability $f$ to submit a news. As such, there will be $f\times N$ news propagating in the network. After a news/story $\alpha$ is submitted (or received) by a user, it will infect each of this user's susceptible neighbors with probability $\beta$. After infecting neighbors, the user will immediately get recovered. All the users who received (or get infected by) $\alpha$ at the end will be recorded. For each user $i$, the set of news/story that she received is denoted as $\Gamma(i)$.

\section{Methods and Metric.}

\subsection{Methods}
In the following, we will describe the method we used to infer the network topology based on the news propagation process. The basic idea is that the news/stories received by users can be used to estimate the similarity between them (nodes). We assume that the nodes with higher similarity are more likely to be connected in networks. Therefore, the obtained similarity $s_{ij}$ can be regarded as the likelihood score $L_{ij}$ for two nodes to have a link, i.e. $L_{ij}=s_{ij}$.

Actually, the similarity $s_{ij}$ is subject to different definition. Here we consider some well-known similarity definitions as follows.

(i) \emph{Common Neighbours (CN)}-By common sense,
two nodes, $i$ and $j$, are more likely to have a link
if they received many same news/stories. The simplest
measure of this neighbourhood overlap is the directed
count, namely
\begin{equation}
s_{ij}=|\Gamma(i)\cap\Gamma(j)|.
\end{equation}

(ii)\emph{Salton Index (SI)}-The Salton index~\cite{salton} is defined as
\begin{equation}
s_{ij}=\frac{|\Gamma(i)\cap\Gamma(j)|}{\sqrt{|\Gamma(i)|\times |\Gamma(j)|}}
\end{equation}
where $|\Gamma(i)|$ the number of news received by user $i$.

(iii) \emph{Jaccard Index (JI)}-This index was proposed by Jaccard over a hundred years ago~\cite{jaccard}, and is defined as
\begin{equation}
s_{ij} =\frac{|\Gamma(i)\cap\Gamma(j)|}{|\Gamma(i)\cup\Gamma(j)|}.
\end{equation}

(iv) \emph{Sorensen Index (SSI)}-This index is used mainly for ecological community data~\cite{sorensen}, and is defined as
\begin{equation}
s_{ij} =\frac{2\times|\Gamma(i)\cap\Gamma(j)|}{|\Gamma(i)|+|\Gamma(j)|}.
\end{equation}

(v) \emph{Hub Promoted Index (HPI)}-This index is proposed for
quantifying the topological overlap of pairs of substrates in metabolic networks~\cite{science2971553}, and is defined as
\begin{equation}
s_{ij}=\frac{|\Gamma(i)\cap\Gamma(j)|}{min\{|\Gamma(i)|,|\Gamma(j)|\}}.
\end{equation}

(vi) \emph{Hub Depressed Index (HDI)}-There is a measure with the opposite effect on
hubs, which is
\begin{equation}
s_{ij} =\frac{|\Gamma(i)\cap\Gamma(j)|}{max\{|\Gamma(i)|,|\Gamma(j)|\}}.
\end{equation}

(vii) \emph{Leicht-Holme-Newman Index (LHN)}-This index assigns
high similarity to node pairs that have many common
neighbours compared to the expected number of such neighbours~\cite{LPN}. It
is defined as
\begin{equation}
s_{ij}=\frac{|\Gamma(i)\cap\Gamma(j)|}{|\Gamma(i)|\times|\Gamma(j)|}.
\end{equation}

(viii) \emph{Resource Allocation Index (RA)}-The similarity between $i$ and
$j$ is defined as the amount of resource $j$ received from
$i$~\cite{EPJB71623}, which is
\begin{equation}
s_{ij} =\sum_{\alpha\in\Gamma(i)\cap\Gamma(j)}\frac{1}{m_{\alpha}}
\end{equation}
where $m_{\alpha}$ is the number of users who finally received news $\alpha$.

As a benchmark, we compare the similarity-based method with the well-known Preferential Attachment (PA) process. The mechanism of preferential attachment has been used to generate evolving scale-free networks, where the probability that a new link is connected to the node $i$ is proportional to $k(i)$~\cite{science286509}. Based on this network growing mechanism, the likelihood score for two nodes to have a link can be calculated as $L_{ij} = |\Gamma(i)|\times|\Gamma(j)|$.

\subsection{Metric}
To measure the accuracy of the method in inferring the network topology, we use the standard metric of the
area under the receiver operating characteristic curve (AUC)~\cite{AUC}. In the network topology inference problem, there are four possible outcomes from the prediction. A true positive (TP) is the prediction of a link that exists in the real network, and if the link doesn't exist in the real network then it is called a false positive (FP). Conversely, a true negative (TN) means that a link that doesn't exist in the real network is not predicted, and a false negative (FN) is the lack of prediction of a link that actually exists in the real network.

To draw the receiver operating characteristic curve (ROC) curve, only the true positive rate (TPR) and false positive rate (FPR) are needed. The TPR defines how many TP occur among all TP and FN samples available during the test. On the other hand, FPR defines how many FP occur among all FP and TN samples available during the test. The ROC curve is created by plotting TPR vs. FPR at various threshold settings. When using normalized units, the area under the ROC curve (AUC) is equal to the probability that a true link has a higher score than a nonexisting link.

In this paper, we use a simple way to calculate AUC. We pick a true link and a nonexisting link in the network and compare their scores. If, among $n$ pairs,
the real link has a higher likelihood score $L_{ij}$ than the nonexisting link $n1$ times and equal score $n2$ times, the AUC value is as follows:
$AUC=(n1+0.5*n2)/n$.
Note that, if links were ranked at random, the AUC value would be equal to $0.5$. By reanalyzing the following results with another accuracy measure, we verify that the performance of the methods is not strongly influenced by the accuracy measure we used. Therefore, we only present the results of AUC in next section.

\begin{figure}
  \center
  \includegraphics[width=12cm]{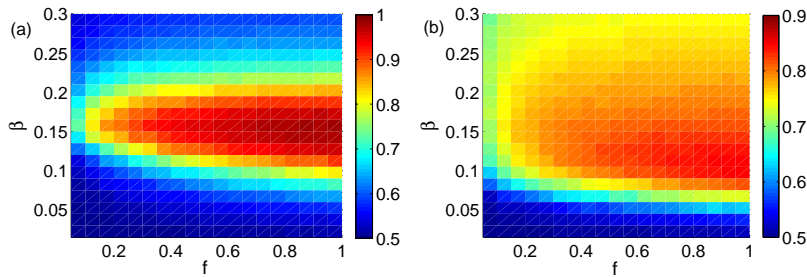}
\caption{(Color online) The $AUC$ in the parameter space ($\beta$, $f$) for (a) WS networks ($N=500$, $p=0.1$, $\langle k\rangle=10$) and (b) BA networks ($N=500$, $\langle k\rangle=10$). The results are averaged over $10$ independent realizations.}
\label{fig1}
\end{figure}

\section{Results}

\subsection{Artificial networks.} We first test our method in two artificial network models: (i) Watts-Strogatz networks (WS)~\cite{Nature393440}, (ii) Barabasi-Albert networks (BA)~\cite{science286509}. When implementing our method, we select the Jaccard similarity definition as an example here. Fig. 1 shows the $AUC$ in the parameter space ($\beta$, $f$) for both WS and BA networks. Actually, both $\beta$ and $f$ control the amount of data we can obtain from the spreading process. If $\beta$ is too small, the news can only propagate several steps and the data for similarity calculation will be limited. If $f$ is small, only a few news are propagating in the network and the obtained similarity matrix will be sparse as well. The first crucial observation in Fig.1 is that the surface of $AUC$ has a pronounced maximum around $\beta=0.15$ in WS networks and $\beta=0.1$ in BA networks for all values of $f$. As discussed above, a small $\beta$ will result in a sparse similarity matrix and eventually lead to a poor $AUC$. In the case of large $\beta$ values, the spreading will cover almost all the network. Consequently, the information of local network structure cannot embed in the spreading results. The optimal $\beta$ is somehow close to the critical infection rate for the spreading coverage~\cite{RMP801275}. Compared to $\beta$, the influences of $f$ on $AUC$ is smaller. Even though $AUC$ keeps increasing with $f$, the increasing speed becomes significantly slower once $f$ is larger than $0.3$.

Next, we move to investigate how the network structure properties influences the inferring accuracy. From Fig. 1, we can already see that $AUC$ in BA networks is lower than that in WS networks, which indicates that it is generally easier to infer the network with homogeneous degree distribution. Furthermore, we study the effect of average degree on the inferring accuracy in detail, with results reported in Fig. 2. Fig.2 (a) and (b) show that as the average degree $\langle k\rangle$ increases, the curve of AUC shifts to the left in both networks. In Fig. 2(c) and (d), we can see that both the optimal $\beta^*$ and maximum $AUC^*$ decrease with $\langle k\rangle$. Interestingly, $\beta^*$ is very stable under different $f$. In WS networks, $\beta^*$ stays almost unchanged when changing $f$. In BA network, $\beta^*$ slightly decreases as $f$ increases.

\begin{figure}
  \center
  \includegraphics[width=12cm]{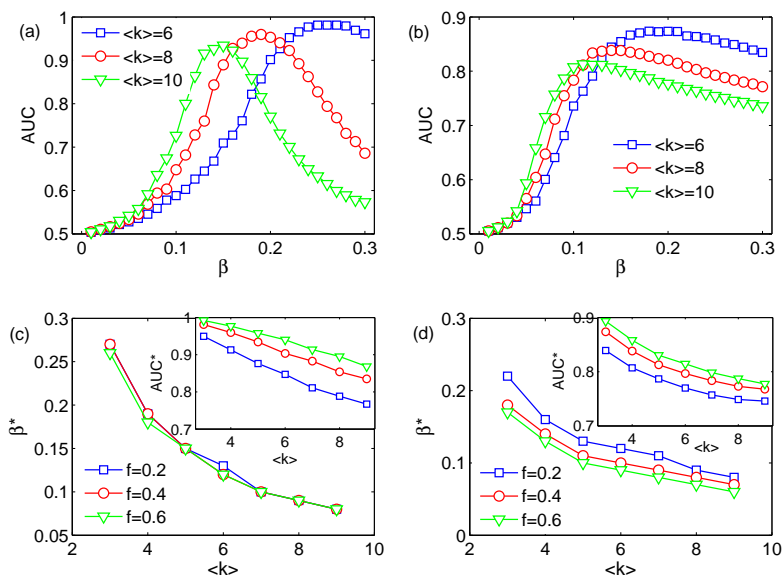}
\caption{(Color online) The dependence of $AUC$ on $\beta$ under different $\langle k\rangle$ in (a) WS networks ($N=500$, $p=0.1$) and (b) BA networks ($N=500$), respectively. (c) and (d) shows the relation between the optimal $\beta^*$ and $\langle k\rangle$ under different $f$ in WS networks and BA networks, respectively. The inset in (c) and (d) are the relation between the maximum $AUC^*$ and $\langle k\rangle$ under different $f$. The results are averaged over $10$ independent realizations.}
\label{fig2}
\end{figure}

We further apply our method on the artificial networks with different size. We present the maximum $AUC^*$ (with respect to optimal $\beta^*$) against $N$ under different $\beta$ in Fig. 3(a) and (b). Interestingly, the inferring accuracy constantly increases with the network size. The curve with the optimal $\beta^*$ enjoys the largest slope ($\beta^*=0.15$ in WS networks and $\beta^*=0.1$ in BA networks). However, the slope slowly becomes smaller as $N$ increases. In Fig. 3(c) and (d), we report the maximum $AUC^*$ against $N$ under different $f$. The results show that $f$ can always improve $AUC^*$.

\begin{figure}
  \center
  \includegraphics[width=12cm]{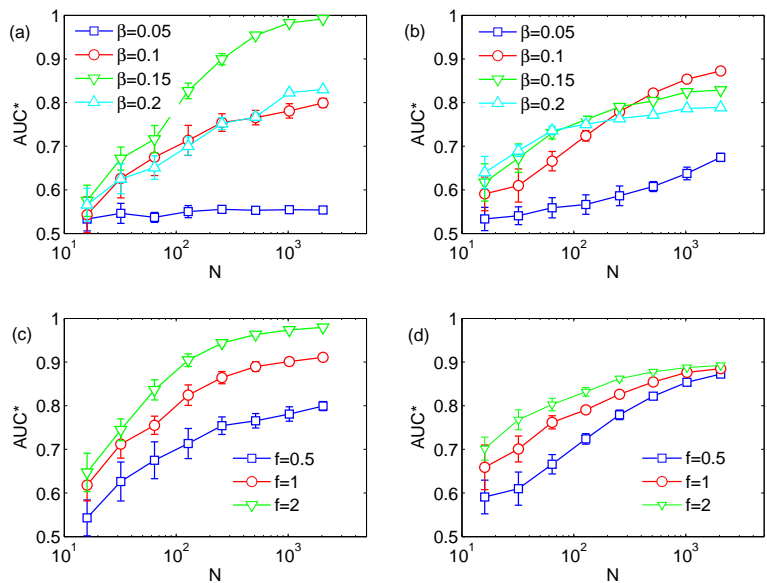}
\caption{(Color online) The maximum $AUC^*$ (with respect to optimal $\beta^*$) against $N$ under different $\beta$ in (a) WS networks ($\langle k\rangle=5$, $p=0.1$) and (b) BA networks ($\langle k\rangle=5$), respectively. (c) and (d) show the maximum $AUC^*$ against $N$ under different $f$ in WS and BA networks, respectively. The results are averaged over $10$ independent realizations.}
\label{fig3}
\end{figure}

\begin{table*}[!t]
\caption{$AUC$ of different similarity definitions in real undirected networks. The parameters are set as $\beta=1/\langle k\rangle$ and $f=0.5$. The similarity with best performance in each network is highlighted in bold font.}
 \begin{tabular}{lccccccccccc}
    \hline
     &$CN$ &$SI$  &$JI$ &$SSI$ &$HPI$ &$HDI$ &$LHN$ &$RA$ &$PA$\\
    \hline
  Dolphins     &$0.8098$ &$0.8088$ &$\textbf{0.8351}$ &$0.8164$ &$0.7836$ &$0.8110$ &$0.7989$ &$0.8200$ &$0.6678$\\
  Word       &$0.8082$ &$0.8109$ &$0.8041$ &$0.8044$ &$0.7774$ &$0.7921$ &$0.6747$ &$\textbf{0.8192}$ &$0.7674$\\
  Jazz     &$0.7918$ &$0.7933$ &$0.7891$ &$0.8007$ &$0.7370$ &$0.7925$ &$0.6876$ &$\textbf{0.8041}$ &$0.7552$\\
  E. coli     &$0.8712$ &$\textbf{0.9022}$ &$0.8944$ &$0.8943$ &$0.8345$ &$0.8918$ &$0.7689$ &$0.8900$ &$0.8302$\\
  USAir     &$0.9086$ &$\textbf{0.9145}$ &$0.9074$ &$0.9066$ &$0.8510$ &$0.8999$ &$0.6524$ &$0.9132$ &$0.8984$\\
  Netsci     &$0.8998$ &$\textbf{0.9186}$ &$0.9183$ &$0.9167$ &$0.9086$ &$0.9148$ &$0.9071$ &$0.9138$ &$0.6672$\\
  Email     &$0.8439$ &$\textbf{0.8758}$ &$0.8676$ &$0.8670$ &$0.8157$ &$0.8554$ &$0.7276$ &$0.8558$ &$0.8131$\\
  TAP      &$0.8691$ &$0.9033$ &$0.9065$ &$\textbf{0.9082}$ &$0.8854$ &$0.9034$ &$0.8903$ &$0.8942$ &$0.7223$\\
  PPI    &$0.8937$ &$0.9345$ &$\textbf{0.9349}$ &$0.9342$ &$0.8613$ &$0.9324$ &$0.8117$ &$0.9124$ &$0.8404$\\
  \hline
 \end{tabular}
\end{table*}

In reality, the infection rate might not be the same in different spreading processes. For example, some news are interesting and thus propagate wider than other news. Besides this, the spreading may only originate from a small region in the network. In the following, we investigate the non-uniform spreading parameters and localized initial condition in the SW and BA models.

In order to model the non-uniform spreading parameters, we modify the spreading process above. Specifically, the infection rate is no longer a constant. After a node is randomly selected as the initial spreader, an infection rate will be set as a random value in the range of [$\beta -\epsilon$, $\beta+\epsilon$]. $\beta$ is the average infection rate and $\epsilon$ is the error magnitude. When $\epsilon=0$, the spreading process reduces to the SIR model we considered before. Once $\epsilon>0$, the infection rate will be different in each spreading process (i.e., each initial spreader selection is corresponding to a different infection rate setting).

We also model the localized initial condition. Instead of selecting the initial spreader from all the nodes in the network, we now consider only the nodes in one specific region as the initial spreader candidates. In practice, we randomly select a node as the seed and calculate the shortest path length from the seed to all the other nodes. The $\eta*N$ nodes with the smallest shortest path length to the seed will form the region for the initial spreader candidates. Clearly, the region is as large as the whole network when $\eta=1$. Once $\eta<1$, the spreading can only originate from a part of the network.

\begin{figure}
  \center
  \includegraphics[width=12cm]{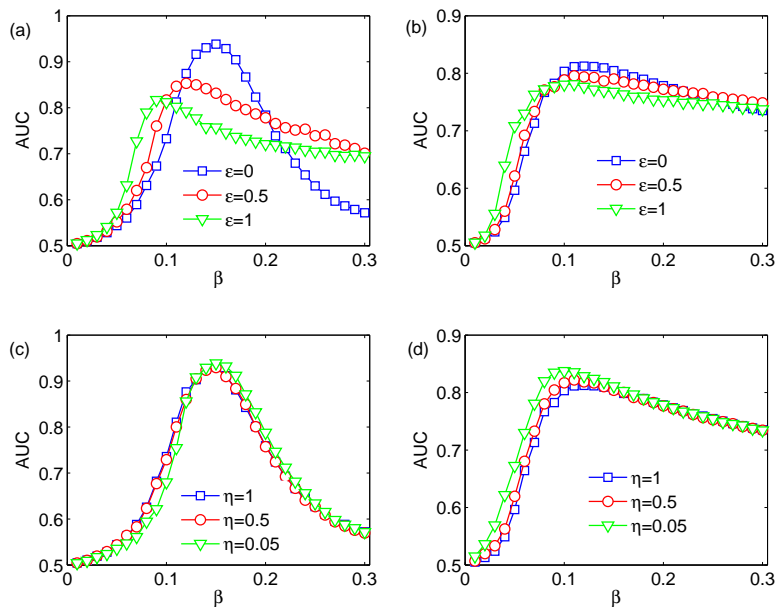}
\caption{(Color online) AUC versus $\beta$ under the non-uniform infection rate setting in (a) SW and (b) BA networks. AUC versus $\beta$ under the localized initial condition in (c) SW and (b) BA networks. In this figure, $f=0.4$. The network parameters are WS ($N=500$, $p=0.1$, $\langle k\rangle= 10$) and BA networks ($N=500$, $\langle k\rangle=10$). The results are averaged over $10$ independent realizations.}
\label{fig3}
\end{figure}

Fig. 4(a) and (b) show the effect of the non-uniform spreading parameters on the inference accuracy. We already discussed that neither small nor large $\beta$ is good for inferring network topology. This is because the similarity matrix is too sparse under small $\beta$ while the similarity between nodes cannot be accurately estimated under large $\beta$ since the viruses cover almost the whole network. The non-uniform spreading parameter setting can increase/decrease some infection rates in spreading. This makes both the small $\beta$ case and large $\beta$ case have some spreading processes with infection rate close to the optimal $\beta^*$, which leads to an improvement in AUC under these $\beta$s. However, the non-uniform spreading parameter setting may significantly lower the maximum AUC and the optimal $\beta$ will be shifted to a smaller value.

Fig. 4(c) and (d) show the effect of localized initial condition on the inference accuracy. Actually, the localized initial condition mainly influences the results under small $\beta$. When $\beta$ is very large, the spreading covers almost the whole network and the spreading results will be independent of the original spreaders. In SW networks, the localized initial condition will lower the accuracy under small $\beta$. This is because a large part of the network have no spreading record to calculate the similarity matrix. Interestingly, the localized initial condition seems to improve the accuracy under small $\beta$ in BA networks. BA networks have some hub nodes which connect to almost all the other nodes in the network and these hub nodes can effectively enhance the local spreading to global level (so that the similarity matrix won't be too sparse). In the local region where the initial spreaders are chosen, the inference accuracy becomes better since more spreading information is available for calculating the similarity.

\subsection{Real undirected networks.} We will validate our method in real undirected networks and all the similarity definitions discussed above will be compared. Both social and nonsocial networks are selected.

The social networks are:  Dolphins (friendship network with $62$ nodes and $159$ links)~\cite{dolphins}, Jazz (musical collaboration network with $198$ nodes and $2742$ links)~\cite{jazz}, Netsci (collaboration network of network scientists with $379$ nodes and $914$ links)~\cite{netcoauthor_word}, Email (email communication network with $1133$ nodes and $5451$ links)~\cite{email}.

The nonsocial networks are: Word (adjacency network in English text with $112$ nodes and $425$ links)~\cite{netcoauthor_word}, E. coli (metabolic network of E. coli with $230$ nodes and $695$ links)~\cite{Ecoli}, USAir (Airline network of USA with $332$ nodes and $2126$ links)~\cite{USAir}, TAP (yeast protein-protein binding network generated by tandem affinity purification experiments, with $1373$ nodes and $6833$ links)~\cite{TAP}, PPI (a protein-protein interaction networks with $2375$ nodes and $11693$ links)~\cite{Nature417399}.

The results in Table 1 show that the similarity based network inferring method can achieve significant higher accuracy than the preferential attachment method. Among the similarity measures we considered, the $SI$, $JI$ and $RA$ generally perform best and are very robust in the performance. We also examine the performance of different similarity metrics in these networks with the non-uniform spreading parameters and localized initial condition. The results show that $SI$, $JI$, $RA$ metrics still generally perform best, and the AUC is not significantly influenced.

\begin{table*}[!t]
\caption{$AUC$ of different similarity definitions in real directed networks. The parameters are set as $\beta=2/\langle k_{out}\rangle$ and $f=0.5$. The similarity with best performance in each network is highlighted in bold font.}
 \begin{tabular}{lcccccccccccc}
         \hline
      &$CN$  &$SI$  &$JI$ &$SSI$ &$HPI$ &$HDI$ &$LHN$ &$RA$ &$AS$ &$PA$\\
        \hline
  Prisoners      &$0.7339$ &$0.8159$ &$0.8133$ &$0.8164$ &$0.7951$ &$0.7987$ &$0.7559$ &$0.7483$ &$\textbf{0.8350}$ &$0.6469$\\
  SM FW    &$0.6643$ &$0.6834$ &$0.6634$ &$0.6543$ &$0.6839$ &$0.6484$ &$0.6127$ &$0.6774$ &$\textbf{0.7635}$ &$0.6111$\\
  LR FW    &$0.7046$ &$0.7135$ &$0.7097$ &$0.7012$ &$0.7052$ &$0.7019$ &$0.7038$ &$0.7102$ &$\textbf{0.7308}$ &$0.6855$\\
  Neural     &$0.7083$ &$0.7076$ &$0.7051$ &$0.7052$ &$0.7049$ &$0.6995$ &$0.6476$ &$0.7209$ &$\textbf{0.7658}$ &$0.6864$\\
  Metabolic  &$0.7043$ &$0.7373$ &$0.7239$ &$0.7246$ &$0.7596$ &$0.7175$ &$0.7027$ &$0.7178$ &$\textbf{0.8031}$ &$0.6542$\\
  PB       &$0.8757$ &$0.8784$ &$0.8761$ &$0.8777$ &$0.8606$ &$0.8722$ &$0.7676$ &$0.8767$ &$\textbf{0.8926}$  &$0.8677$\\
          \hline
 \end{tabular}
\end{table*}

\subsection{Real directed networks.} Actually, our method can be easily extended to directed networks. However, all the similarity measures discussed above are symmetric (i.e. $s_{ij}=s_{ji}$). It implies that if a directed link exists, the link in the other directed will exist as well. This will largely lower the accuracy. To solve the problem, we proposed an asymmetric similarity (AS) measure for inferring the network topology in directed networks. Mathematically, it can be expressed as
\begin{equation}
s_{ij} =\frac{|\Gamma(i)\cap\Gamma(j)|}{|\Gamma(i)|}.
\end{equation}
A large $s_{ij}$ indicates that $j$ received most of the news/stories passing through $i$. Therefore, it is more likely to have a directed link from $i$ to $j$.

We considered several real directed network to validate our method. The networks include Prisoners (friendship network between prisoners with $67$ nodes and $182$ links)~\cite{Prision}, SM FW (food web network in St. Mark area with $54$ nodes and $356$ links)~\cite{FW}, LR FW (food web network in little rock area with $183$ nodes and $2494$ links)~\cite{FW}, Neural (the neural network of C. elegans with $297$ nodes and $2359$ links)~\cite{Celegans}, Metabolic (the metabolic network of C. elegans with $453$ nodes and $2040$ links)~\cite{FW}, PB (the hyper link between the blogs of politicians with $1222$ nodes and $19090$ links)~\cite{PB}. Again, we observe that the inferring accuracy of similarity-based method is higher than the Preferential attachment method. Interestingly, the $AS$ performs best among all the similarity measures. The results indicate that the asymmetric feature is crucial for inferring network topology in directed networks.

Like in undirected networks, we examine the performance of different similarity metrics in directed network with the non-uniform spreading parameters and localized initial condition. We observe that the accuracy is largely lowered. Generally speaking, the virus/information is more difficult to propagate in these networks due to the directionality of the links. Therefore, the virus is very likely to stay in the local region under the localized initial condition, which results in a very sparse similarity matrix for inferring network topology and thus a much lower AUC. The phenomenon is even more serious in some acyclic networks (such as SM food web and LR food web).

\section{Conclusion.} To summarize, we propose a method to infer the network topology based on the spreading process on networks. Specifically, the similarity between nodes are estimated by the information/virus that nodes received, and the nodes with the highest similarity are assumed to be connected. We tested our method in classic artificial network models and find that our method enjoys high inferring accuracy. Moreover, we find that the infection rate in the spreading process significantly affects the inferring results and there is an optimal infection rate for each network. The findings are confirmed in many real networks. Finally, the method is extended to directed networks. We proposed a new similarity measure, which is shown to perform better than other well-known similarity measures in directed networks.

We remark that many extensions can be made in this direction. For example, the inferring accuracy can be further improved if the time information of the spreading is known (i.e., at what time the nodes receive the virus). In addition, it is interesting and important to design an more efficient method for the cases where only partial information of the spreading can be obtained.

\section*{Acknowledgement.} The author would like to thank an anonymous referee for the comments to improve the paper. The author acknowledges the support from China Scholarship Council.

\end{document}